\documentclass[twocolumn,twoside,slac_two]{revtex4}
\usepackage{graphicx}
\usepackage{fancyhdr}
\begin{document}

\title{Speeding up simulations of relativistic systems using an optimal boosted frame}

%

\author{J.-L. Vay, W. M. Fawley, C. G. Geddes, E. Cormier-Michel}
\affiliation{Lawrence Berkeley National Laboratory, Berkeley, CA, USA}
\author{D. P. Grote} 
\affiliation{Lawrence Berkeley National Laboratory, Livermore, CA, USA}

\begin{abstract}
It can be computationally advantageous to perform 
computer simulations in a Lorentz boosted frame for a certain class of 
systems. However, even if the computer model relies on a covariant set of 
equations, it has been pointed out that algorithmic difficulties related to 
discretization errors may have to be overcome in order to take full advantage 
of the potential speedup. We summarize the findings, the 
difficulties and their solutions, and show that the technique enables simulations 
important to several areas of accelerator physics that are otherwise problematic, 
including self-consistent modeling in three-dimensions of laser wakefield accelerator stages at energies of 10 GeV and above.
\end{abstract}

\maketitle

\thispagestyle{fancy}


\section{Introduction}
In \cite{VayPRL07}, we have shown that the ratio of longest to shortest space
and time scales of a system of two or more components crossing at relativistic
velocities is not invariant under a Lorentz transformation. This implies the
existence of an ``optimum''  frame of reference minimizing a measure of the ratio of space
and time scales. Since the number of computer operations (e.g., time steps),
for simulations based on formulations from first principles, is proportional
to the ratio of the longest to shortest time scale of interest, it follows that
such simulations will eventually have different computer runtimes, yet
equivalent accuracy, depending solely upon the choice of frame of
reference. The scaling of theoretical speedup was derived for a generic case of
two crossing identical rigid particle beams, and for three particle
acceleration related problems: particle beams interacting with electron clouds
\cite{ecloud}, free electron lasers (FEL) \cite{fel}, and laser-plasma accelerators
(LWFA) \cite{lwfa}. For all the cases considered, it was found that the ratio of space
and time scales varied as $\gamma^2$ for a range of $\gamma$, the relativistic factor 
of the frame of reference relative to the optimum frame. For systems
involving phenomena (e.g., particle beams, plasma waves, laser light in
plasmas) propagating at large $\gamma$, demonstrated speedup of simulations
being performed in an optimum boosted frame can reach several orders of
magnitude, as compared to the same simulation being performed in the laboratory frame. 

We summarize the difficulties and limitations of the method, the
solutions that were developed, and the simulations that we have performed to date. 
We show that the technique enables simulations 
important to several areas of accelerator physics that are otherwise problematic. 
For the first time, it allows for direct self-consistent simulations of laser wakefield accelerator stages at 10 GeV and 
beyond using current supercomputers in a few hours, while the same calculations in 
the laboratory frame would take years using similar resources and are thus impractical.
It also allows simulations of electron cloud effects in high energy physics accelerators (modeled so 
far with codes based on quasistatic approximations) using more standard Particle-In-Cell methods. This renders these 
types of simulations  accessible to a wider range of existing computer codes, alleviates 
the added complication due to pipelining when parallelizing a quasistatic code, and removes 
the approximations of the quasistatic method which may not be applicable in some situations.
For free electron lasers, the new technique offers the possibility of calculating self-consistently 
configurations that are not accessible with standard FEL codes due to the 
limitations of the approximations that they are based on. Finally, the method may offer a unique way 
of calculating self-consistently, and in three-dimensions, coherent synchrotron radiation effects which 
are of great importance in several current and future accelerators.


\section{Difficulties}
Even if the fundamental electrodynamics and particles equations are written in a covariant form, the numerical algorithms that are derived from them may not retain this property and special techniques have been developed to allow simulations in boosted frames. As an example, we considered in  \cite{VayPOP08} an isolated beam 
propagating in the laboratory frame at relativistic velocity. 
When applying the effect of the beam field on itself using the Newton-Lorentz equation of motion, the contribution from the radial electric field is largely canceled by the contribution from the azimuthal magnetic field. However, we showed that the so-called `Boris particle pusher' \cite{BorisICNSP} (which is widely used in PIC codes), does make an approximation in the calculation of the Lorentz force which leads to an inexact cancellation of the electric component by the magnetic component. The magnitude of the error grows with the beam relativistic factor and in practice, it is unacceptably large for simulations of ultra-relativistic charged beams, where the cancellation needs to be nearly complete. 
The issue was resolved by changing the form of the Lorentz force term in the Boris pusher, and solving analytically the resulting implicit system of equations (see \cite{VayPOP08} for details).

An additional practical complication of numerical simulation in a boosted frame
is that inputs and outputs are often specified (or desired) in the laboratory
frame. For example, in LWFA simulations, laser and plasma parameters have to be
transformed from the laboratory to the new relativistic boosted frame, so that the
electromagnetic waves will be Doppler-shifted, and the background plasma, with
higher density, is now drifting. In the PIC code Warp \cite{Warp}, the laser is
injected at a plane that is fixed in the laboratory frame and drifting in the
boosted frame.
Likewise, the initial phase-space distribution of a particle beam is
generally known in the laboratory. For calculations in boosted frames of large
$\gamma$, deriving the initial beam conditions at a given time can be easy if the 
initial conditions are simple (e.g., initial Gaussian beam in vacuum), or more difficult
and/or computationally costly if injecting the beam in a particle
accelerator for example, where its longitudinal extent in the boosted frame can cover
several lattice periods. In order to circumvent this difficulty, a procedure
was implemented in Warp which injects the beam through a transverse plane that
is fixed in the laboratory, but drifting in the boosted frame, similarly to the
laser injection method. Due to long range space charge forces, it is still
necessary to provide a reasonable estimate of the beam distribution near the
injection plane; this is accomplished by the use of ``frozen" drifting
macroparticles.

After the relativistic PIC algorithm evolves the system in the boosted frame,
the results must be transformed back to the laboratory frame. We have found it convenient in Warp to record
quantities at a number of regularly spaced ``stations", immobile in the
laboratory frame, at a succession of discrete times, for both detailed time histories 
and laboratory time-averages. Since the space-time locations of the diagnostic grids in the laboratory frame generally do not coincide with the space-time positions of the macroparticles and grid nodes used for the calculation in a boosted frame, some interpolation is performed during the data gathering process. 
Finally, in simulations of laser-plasma acceleration stages (see below), we observed a short wavelength instability with a growth rate that rises with the velocity of the boosted frame and the inverse of the grid resolution, which we have controlled through the use of low dispersion electromagnetic solvers \cite{KarkICAP06} and low-pass digital filtering. The details of the instability and its cures will be detailed in a future paper 
\cite{Vayprep09}.

Together with mitigation of numerical artifacts as just described, these techniques allow simulations using boosted frames, with orders of magnitude speedup over the same simulations performed using a laboratory frame, as shown below. Additional details of the input and output procedures can be found in \cite{VayPAC09}.

\section{Examples of application}

\subsection{Laser wakefield acceleration}
Laser driven plasma waves offer orders of magnitude increases in accelerating
gradient over standard accelerating structures (which are limited by electrical
breakdown), thus holding the promise of much shorter particle
accelerators. Yet, computer modeling of the wake formation and beam
acceleration requires fully kinetic methods and large computational resources
due to the wide range of space and time scales involved \cite{RefCameron}. For
example, modeling 10 GeV stages for the LOASIS (LBNL) BELLA proposal
\cite{BELLA} in one-dimension demanded as many as 5,000 processor hours on a
NERSC supercomputer \cite{BruhwilerAAC08}. As discussed in \cite{VayPRL07}, the
range of scales can be greatly reduced if one adopts the common assumption that
the backward-emitted radiation can be neglected, enabling, for the first time,
the full-PIC simulation of the next generation of laser systems.

Warp simulations at plasma density $n_e=10^{19}$ cm$^{-3}$ were performed in
2-1/2D and 3D using reference frames
moving anywhere between $\gamma_f=1$ (laboratory frame) and $10$.  These
simulations are scaled replicas of $10$ GeV stages that would operate at actual
densities of $10^{17}$ cm$^{-3}$ \cite{CormierAAC08,GeddesPAC09} and allow
short run times to permit effective benchmarking between the algorithms. Agreement within a few percent was observed on
the beam peak energy and average energy between calculations in all frames, showing that the boosted frame simulations
gain speed without sacrificing accuracy. A speedup of 100 was 
measured between the calculation in the frame at $\gamma=10$ and the
calculation in the laboratory frame. 

The boosted frame model was then used to conduct full scale simulations of 10 GeV stages at plasma densities of  $10^{17}$ cm$^{-3}$  in 2-1/2D and 3D simulations.  Simulations at $10^{18}$ cm$^{-3}$ were also conducted to establish scaling.  Relativistic factors of the boosted frame were 130 and 40 respectively, i.e. close to the relativistic factor associated with the wake velocity in the laboratory frame.
The 3D run at full scale took almost 4 hours with $\gamma_f$ = 130 using 512 cores on the cluster Lawrencium at LBNL. This provided direct simulation of next generation experiments and possible laser-plasma collider stages. Good agreement with the scaled energy gain was obtained.  Such simulations are impractical in the laboratory frame, with projected time of 15 years on the same resources using the $2\gamma^2_f$ formula for the estimated speedup, and scaling from standard PIC runs.

\subsection{Electron cloud driven instabilities}
Several existing and planned future particle accelerators have limitations due
to the electron cloud instability that may negatively impact the beam quality
and in some cases even lead to severe beam loss. A calculation of electron
cloud driven instability \cite{ecloud} for an ultra-relativistic beam was
performed with the Warp code framework in (a) standard PIC mode using the new
particle pusher in a Lorentz boosted frame; (b) in quasistatic mode
\cite{Quickpic} using linear maps to push beam particles into the
accelerator lattice. The two runs were in good agreement and completed using
similar computer resources and runtimes. The speedup factor of the PIC boosted
frame calculation compared to a PIC calculation in the laboratory frame was
estimated at 500. For many calculations of electron cloud instability, the
boosted frame approach may not resolve any additional physics not included in
the quasistatic approach. 
We note, however, that the quasistatic method requires significant special coding 
to take into account eventual 
longitudinal motion of electrons \cite{Quickpic}, as well as a special parallelization scheme 
\cite{FengJCP09} for parallelization along the axis of beam propagation, which
are not standard to PIC codes. By contrast, the boosted frame method 
includes naturally the longitudinal dynamics and requires
more modest modifications to an existing standard PIC code or framework
(none for parallelization, if the PIC code is already parallel).

\subsection{Free electron lasers}
In a short wavelength free-electron laser, a high energy electron beam
interacts with a static magnetic undulator. In the optimal boost frame with
Lorentz factor $\gamma$, the down-shifted FEL radiation and up-shifted
undulator have identical wavelengths and the number of required time-steps
(presuming the Courant condition applies) decreases by a factor of $2\gamma^2$
for fully electromagnetic simulations. Examples of boosted-frame simulations
have been compared \cite{FawleyPAC09} to results obtained with the
eikonal (i.e, SVEA) and wiggler-period averaged code Ginger \cite{Ginger}. It
was concluded that if the necessary FEL physics can be studied with an eikonal
code, it will run much faster than a full electromagnetic code in whatever frame. However,
if there are important physical phenomena that cannot be resolved properly by
an eikonal code, a boosted-frame electromagnetic code is a very attractive
alternative to a brute force full electromagnetic calculation in the laboratory frame.

\subsection{Coherent synchrotron radiation}
Another application for which the Lorentz-boosted frame method might be useful
is that of modeling coherent synchrotron radiation (CSR) \cite{CSR} emitted by high current,
high brightness relativistic electron beams. 
Because full scale electromagnetic simulation of CSR in the laboratory frame
is difficult due to the wide range of scales (chicane lengths of order meters,
radiation wavelengths of orders microns), in order to make the calculation
tractable most CSR simulation codes
apply simplifications such as ignoring transverse variation of CSR
across the electron beam. 
We have begun preliminary work of simulating CSR emission with the boosted
frame method with Warp, examining the behavior of a high current, short electron beam transiting
a simple dipole magnet. 
Our early results show that upon
exit from the undulator the electron beam shows the characteristic energy loss
variation with longitudinal position that one expects from previous theoretical
analyses of CSR. Further studies are currently underway.
\section{Conclusion}
The non-invariance of the range of scales of a physical system implies that the
computational cost of a certain class of computer simulations depends strongly
on the choice of the simulation frame of reference. Algorithmic
difficulties arise due to the loss of covariance upon discretization of the
Maxwell-Vlasov system of equations, and the need to transform input/output data
between the laboratory frame and the Lorentz boosted frame. So far, the difficulties that have arisen 
have been overcome and no ``show-stopper'' has been identified at this time. First
principles simulations in boosted frames have been performed successfully with the
code Warp in application to laser wakefield acceleration, electron cloud driven
instabilities and free electron lasers, with
speedups ranging between a few and several orders of magnitude. 
Our recent progress show that first principles modeling in a Lorentz boosted
frame is a viable alternative or complement to using reduced descriptions like
the quasistatic \cite{Quickpic} or eikonal \cite{Ginger}
approximations, or performing simulations with scaled parameters \cite{GeddesPAC09}, 
and in many cases includes physics that is not accessible to the other descriptions. 
This includes direct three-dimensional simulations of laser wakefield 
accelerator stages at 10 GeV and beyond, electron cloud effects in high energy physics
accelerators, physics that is inaccessible to standard free electron lasers codes, and coherent 
synchrotron radiation.

\begin{acknowledgments}
We are thankful to D. L. Bruhwiler, J. R. Cary, B. Cowan, E. Esarey, M. A. Furman, C. Huang, S. F. Martins, W. B. Mori, B. A. Shadwick, C. B. Schroeder and M. Venturini for insightful discussions.

This Work was supported by US-DOE Contracts DE-AC02-05CH11231 and DE-AC52-07NA27344, US-LHC program LARP, and US-DOE SciDAC program ComPASS. It Used resources of NERSC, supported by US-DOE Contract DE-AC02-05CH11231.
\end{acknowledgments}

\bigskip 

\end{document}